\documentclass[amsmath,twocolumn]{aastex701}
\usepackage{acro}[=v2]
\acsetup{display-foreign = false}
\acsetup{foreign-format = {\scshape} }
\ExplSyntaxOn
\prop_new:N \l__acro_foreign_format_prop
\ExplSyntaxOff
\usepackage{new_commands}
\date{\today}

\begin{document}

\title{The steep redshift evolution of the hierarchical binary black hole merger rate may cause the $z$-$\chi_{\rm eff}$ correlation}

\author[0000-0002-6121-0285]{Amanda M. Farah}
\email{afarah@cita.utoronto.ca}
\CITA
\author[0000-0002-4103-0666]{Aditya Vijaykumar}
\email{aditya@utoronto.ca}
\CITA
\author[0000-0002-1980-5293]{Maya Fishbach}
\email{fishbach@cita.utoronto.ca}
\CITA

\begin{abstract}
There is growing evidence from gravitational-wave observations that some merging black holes are created from previous mergers.
Using the prediction that these hierarchically merged black holes have dimensionless spin magnitudes of $\chi \approx 0.69$, we identify a subpopulation in the gravitational-wave data consistent with a hierarchical-merger origin in dense star clusters.
This subpopulation's primary mass distribution peaks at \result{$17.0^{+18.3}_{-4.4}\Msun$}, which is approximately twice as large as its secondary mass distribution's mode (\result{$10.5^{+29.7}_{-4.7}\Msun$}), and its spin tilt distribution is consistent with isotropy.
Our inferred secondary mass distributions imply that isolated binary evolution may still be needed to explain the entirety of the $9\,\mathrm{M}_{\odot}$ peak.
Surprisingly, we find that the rate of hierarchical mergers may evolve more steeply with redshift than the rest of the population (\result{$98.0\%$} credibility): the fraction of all binary black holes that are hierarchically formed at $z=0.1$ is \result{$0.03^{+0.05}_{-0.02}$}, compared to \result{$0.09^{+0.11}_{-0.07}$} at $z=1$.
This provides an explanation for the previously discovered broadening of the effective spin distribution with redshift.
Our results have implications for star cluster formation histories, as they suggest the potential existence of a high-redshift population of massive, compact clusters.
\end{abstract}


\section{Introduction} 
Star clusters are foundational to much of our understanding of galactic and stellar astrophysics, as they play a large role in galactic structure and dynamics \citep{1979ARA&A..17..241H}, significantly contribute to the chemical evolution of their galaxies \citep{2002ApJ...571..830S}, and can be used to trace galaxy assembly \citep{2006ARA&A..44..193B}.
Crucial but uncertain features of the star cluster population, such as its mass function and its evolution with redshift \citep{2018RSPSA.47470616F}, can be probed by the stellar-mass \acp{BBH} observed by the  LIGO~\citep{2015CQGra..32g4001L,2025PhRvD.111f2002C,2025CQGra..42h5016S}, Virgo~\citep{2015CQGra..32b4001A} and KAGRA~\citep{2021PTEP.2021eA101A} (LVK) gravitational-wave detectors \citep{2023MNRAS.522.5546F}.
However, this requires knowing which portion of the \ac{BBH} population is produced by clusters.

The clearest population-level signature of \ac{BBH} formation in star clusters is the existence of hierarchical mergers \citep[e.g.][]{2021NatAs...5..749G}.
These include one or more \ac{2G} remnants of previous, \ac{1G} black hole mergers.
The number of mergers that include a \ac{2G} \ac{BH} strongly informs the total rate of cluster-origin \acp{BBH}.
In this Letter, we investigate the population properties of the hierarchically merged \acp{BBH}, which can be used to inform the redshift and metallicity distributions of star clusters. 
\begin{figure*}[ht!]
    \centering
    \includegraphics[width=\linewidth]{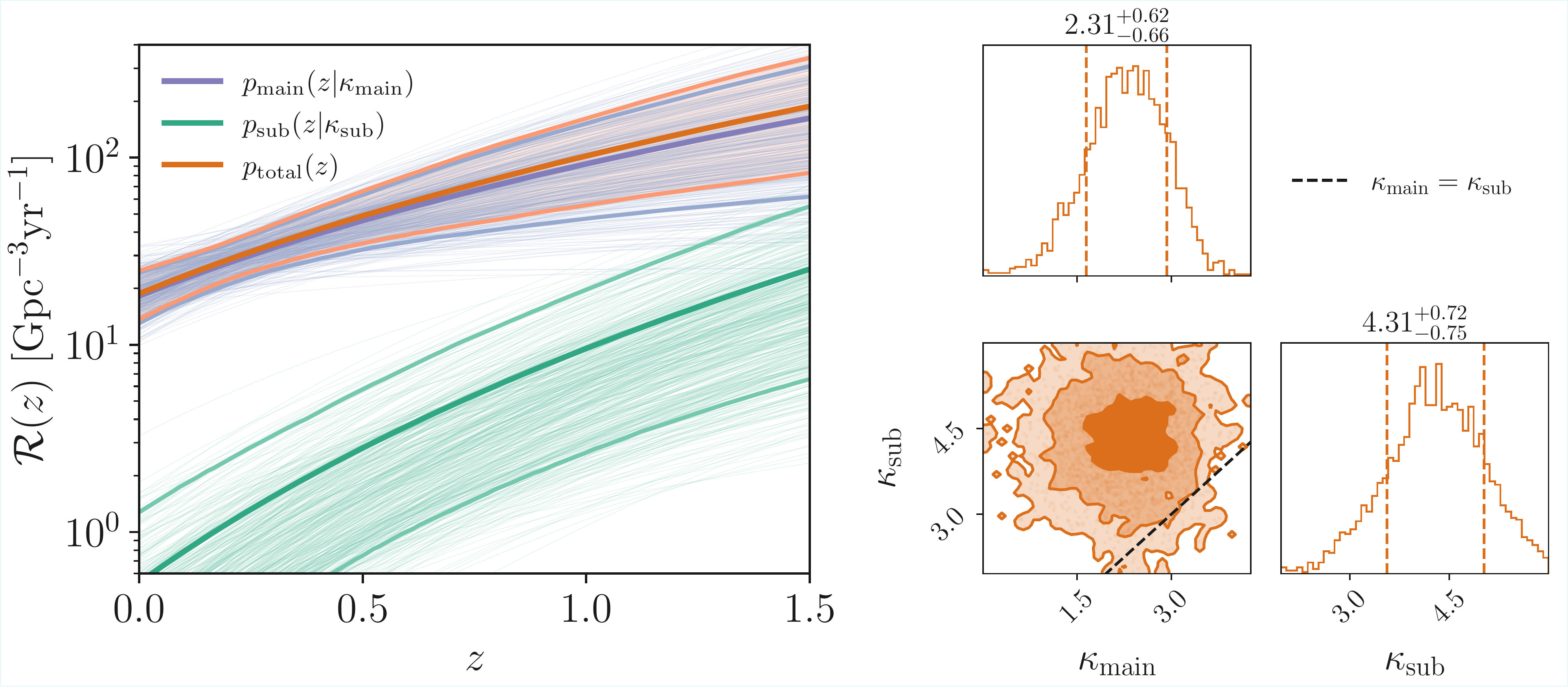}
    \caption{The inferred merger rate evolution with redshift of our subpopulation of $\chi_1\approx0.7$ \acp{BBH} ($p_{\rm sub}(z|\kappa_{\rm sub})$, green in left panel) compared to that of the rest of the population ($p_{\rm main}(z|\kappa_{\rm main})$, violet in left panel).
    Individual draws from the main and subpopulation are shown as thin lines, shaded bands show the $90\%$ credible interval, and dark lines show the mean.
    As shown by the inferred power law indices for the main and subpopulation (right panel), the spinning subpopulation has a steeper redshift distribution than the main population, causing a higher contribution to the overall rate (orange lines in left panel) at high redshift than at low redshift. 
    This is contrary to typical expectations of hierarchical mergers in star clusters, which necessitate slightly longer delay times for hierarchical mergers than the first-generation mergers that precede them.
    The existence of a rapidly-spinning subpopulation with a steep redshift evolution provides a natural explanation of the previously identified correlation between effective spin and redshift in the overall \ac{BBH} population. 
    }
    \label{fig:redshift}
\end{figure*}

Beyond their ability to  identify which portion of the BBH population is sourced from clusters, hierarchical mergers are informative in their own right.
Firstly, they are rarely produced in clusters with low stellar mass, as the recoil kicks that \ac{2G} black holes receive from anisotropic emission of gravitational waves are typically larger than the clusters' escape velocity.
Thus, \ac{2G} black holes are often ejected from star clusters and only a small fraction remain to undergo another merger ~\citep{1983MNRAS.203.1049F, 2000ApJ...528L..17P, 2004ApJ...607L...5F, 2007PhRvL..98i1101G, 2010PhRvD..81h4023L, 2018PhRvD..97j4049G, 2021ApJ...918L..31M}. 
The mass and spin of the \ac{2G} object in a hierarchically merged system therefore provide a lower limit on the escape velocity of its host and insight into the system's environment.

Secondly, hierarchical mergers have clear population predictions that allow them to be straightforwardly distinguished from the larger \ac{BBH} population \citep{2017ApJ...840L..24F, 2017PhRvD..95l4046G}.
An ensemble of \acp{BH} that have resulted from previous mergers will have a spin magnitude distribution peaked near $\chi=0.69$ \citep{2008ApJ...684..822B, 2016ApJ...825L..19H,2019PhRvD.100d3027R,2025ApJ...987..146B}, regardless of how many previous mergers occurred \citep{2017ApJ...840L..24F}.
Additionally, \ac{2G}+\ac{1G} mergers will have a mass ratio distribution peaked near $\approx1/2$ \citep{2017PhRvD..95l4046G,2019PhRvD.100d3027R,2020ApJ...900..177K}, and \acp{BBH} formed in gas-free star clusters should have an isotropic spin tilt distribution (but cf. \citealt{2025ApJ...983L...9K}). 
However, mass ratio and spin orientation distributions differ for 2G+1G mergers formed in the disks of active galactic nuclei \citep{2020ApJ...899...26T, 2020MNRAS.494.1203M,2024A&A...685A..51V,2025arXiv250815337V,2025arXiv250923897L}.

The simplicity and robustness of these predictions contrasts with most other formation pathways whose population predictions suffer from large theoretical uncertainties. 
Hierarchical mergers may therefore be the first subpopulation to be confidently identified, even if they make up a small fraction of the total population.
Indeed, evidence for a hierarchically merged subpopulation in the \ac{GW} data is growing.
In addition to hints that hierarchical mergers are filling the pair-instability mass gap in the \ac{BBH} primary mass distribution \citep{2024arXiv240601679P,2025PhRvL.134a1401A,2025arXiv250904151T,2025arXiv250909123A}, the two \ac{GW} detections recently released by the LVK with primary spin magnitudes of $\approx 0.7$ \citep{2025ApJ...993L..21A} have pointed to hierarchical mergers across the \ac{BBH} mass spectrum \citep{2025arXiv251105316T, 2025arXiv251203152B}, a possibility first pointed out by \citet{2021ApJ...913L..19T}.

In this work, we isolate the hierarchical mergers in the latest \ac{GW} data using a phenomenological model that takes advantage of the most straightforward and robust population prediction for hierarchical mergers: their primary spin magnitude distribution.
We then investigate the population properties of this hierarchically merged subpopulation and provide insight into their cluster environments and the population of \ac{1G} \acp{BH} in star clusters (via the 1G secondaries of 2G+1G mergers).
The redshift distribution of our identified hierarchically merged subpopulation is likely steeper than that of the rest of the population (Fig.~\ref{fig:redshift}).
This is in contrast to na\"ive expectations from cluster dynamics, in which hierarchical mergers occur at similar (or slightly later) times in a cluster's history since they necessarily occur after the first \ac{1G} mergers.
This finding therefore either implies that the subpopulation of $\chi_1\approx0.69$ objects is not formed hierarchically, or that the hierarchical mergers are coming from a distinct population of more distant star clusters than produced our observed first-generation mergers.
Regardless, the fact that a highly-spinning subpopulation has a steeper redshift evolution than the rest of the \acp{BBH} offers a potential explanation to the previously discovered broadening of the effective spin distribution with redshift.
In a companion paper \citep{2026arXiv260103457V}, we show that this subpopulation can also explain the effective spin distribution's narrowing with mass ratio \citep{2021ApJ...922L...5C,2025arXiv250818083T}.

The low-mass end of our inferred 1G mass distribution differs between the 1G+1G and 2G+1G populations by more than would be expected from the simplest mass segregation models.
This suggests that other formation channels, such as isolated binary evolution, contribute to the low-mass 1G+1G population.

This paper is organized as follows.
Section~\ref{sec:model} describes the model we use to isolate a potential subpopulation of \ac{2G}+\ac{1G} \acp{BBH}, and we present the evidence for such a subpopulation in Section~\ref{sec:xi}.
Section~\ref{sec:redshift} describes our findings related to the redshift evolution of this subpopulation, using it to provide measurements of the relative fraction of cluster-origin \acp{BBH} in the overall population. 
Our inferred mass distributions are shown in Section~\ref{sec:mass}.
Section~\ref{sec:correlation} interprets our results in the context of previously identified correlations in the population, including correlations between primary mass and spin and correlations between spin and redshift, and Section~\ref{sec:discussion} discusses their implications for cluster properties and the contribution of the isolated binary channel to the overall BBH merger rate.
The appendices provide additional results, such as the distributions of spin magnitudes, spin orientations, and mass ratios for the subpopulation and main population.

\section{Phenomenological population model}
\label{sec:model}
We construct a phenomenological mixture model to search for a subpopulation, $S$, of 2G+1G hierarchical mergers in the GW data and disentangle it from the main population, $M$.
The subpopulation is defined by its primary spin magnitude distribution, which is peaked at $\chi_1=0.69$.
We emphasize that although this is a strong prior choice designed to isolate a population of 2G+1G mergers, \emph{we do not enforce the existence of this subpopulation}.
The subpopulation's other dimensions (i.e. its mass, spin orientation, and redshift distributions) are fit separately from the main population.
The mixture model therefore has the form
\begin{widetext}
\begin{equation}
\begin{aligned}
    p(m_1,\cos{\theta_1},\chi_1,m_2,\cos{\theta_2},\chi_2, z | \Lambda) &= \xi p_S(m_1, m_2, \cos{\theta_1},\cos{\theta_2}, z|\lambda_S)\mathcal{N}(\chi_1|\mu^{\chi_1}_S,\sigma^{\chi_1}_S) \mathcal{N}(\chi_2|\mu^{\chi_2}_S,\sigma^{\chi_2}_S) \\
    &+ (1-\xi) p_M(m_1, m_2, \cos{\theta_1},\cos{\theta_2}, z |\lambda_M)\mathcal{N}(\chi_1|\mu^{\chi},\sigma^{\chi}) \mathcal{N}(\chi_2|\mu^{\chi},\sigma^{\chi}) ,
\end{aligned}
\end{equation}
\end{widetext}
where $\xi$ is the fraction of systems in $S$.
Here, $m_i$, $\theta_i$, and $\chi_i$ are the component masses, spin tilt angles, and spin magnitudes, respectively.
$\Lambda$ is the set of all hyperparameters,  $\lambda_j$ are the sets of hyperparameters that govern the distribution of masses and spin tilts for both subpopulations, and $j=\{S,M\}$.
Gaussian distributions with mean $\mu$, standard deviation $\sigma$ and truncated in the range $[0,1]$ are represented by $\mathcal{N}(\cdot|\mu,\sigma)$.
Both populations use a pairing function formalism for the mass distribution, allowing us to fit for separate component mass distributions as well as their mass-ratio dependent pairing \citep{2020ApJ...891L..27F}.
We describe the forms of the main and subpopulation in more detail in Appendix~\ref{ap:model}.

Using a hierarchical Bayesian analysis~\citep{2004AIPC..735..195L,2019PASA...36...10T,2022hgwa.bookE..45V}, we fit this model to all BBHs in \ac{GWTC-4.0}, which includes BBHs in all previous catalogs. 
We additionally include GW241110 and GW241011 \citep{2025ApJ...993L..21A}, two events that have been recently published by the LVK because their spins and mass ratios are consistent with a hierarchical merger origin. 
However, excluding these events does not noticeably change our findings.
We omit GW231123 ~\citep{2025ApJ...993L..25A} as its large secondary mass (above $100\,M_\odot$) appears to be an outlier with respect to our inferred secondary mass distributions.
Including this event does not qualitatively affect our redshift or spin inferences, though it slightly reduces the statistical significance of the result shown in Fig.~\ref{fig:redshift}, and it visibly changes the subpopulation's secondary mass distribution. 
Appendix~\ref{ap:231123} further discusses the effects of GW231123 on our inferred mass distributions.

\section{Identification of a $\chi_1\approx0.7$ subpopulation}
\label{sec:xi}
We find \result{strong} evidence for a subpopulation of $\chi_1=0.69$ systems.
The fraction of systems in this subpopulation, $\xi$, is larger than \result{$0.01$} at the \result{$99.9\%$} level, representing a Bayes factor of \result{$49.3$}. 
This subpopulation has spin orientation, mass ratio, and component mass distributions characteristic of hierarchical mergers in star clusters (see Appendices~\ref{sec:spin tilt}, \ref{sec:mass ratio}, and Section~\ref{sec:mass}, respectively).

Marginalized over the range $z=[0,1.9]$, our inferred value of \result{$\xi=0.13^{+0.16}_{-0.10}$} is roughly consistent with previous studies that also identify a 2G+1G-like subpopulation that makes up roughly $12-20\%$ of the population\footnote{\citet{2025PhRvL.134a1401A} estimate that $20\%$ of all mergers are 2G+1G, and \citet{2025arXiv251105316T} estimate the rate of 2G+1G mergers to be $2.2^{+1.9}_{-1.2}\igpcyr$, which corresponds to $\approx12\%$ of the total BBH merger rate inferred by \citet{2025arXiv250818083T}, albeit with a different population model.
\citet{2024arXiv240601679P} find that $\approx2$--$5\%$ of the overall population belong to a $\chi\approx0.7$ subpopulation, which is lower than that inferred in this work, potentially caused by differing inferred redshift distributions.
Additionally, \citet{2021ApJ...915L..35K} find $\rate_{\rm 2G+1G}/\rate_{\rm 1G+1G} < 0.12$ when analyzing GWTC-2, which is in agreement with our posterior on this value, which peaks at \result{$0.07$}.} \citep[e.g.][]{2021ApJ...915L..35K,2022ApJ...941L..39W, 2022PhRvD.106j3013M,2024arXiv240601679P,2025PhRvL.134a1401A,2025arXiv250904151T,2025arXiv251105316T,2025arXiv251203152B,2026arXiv260103457V}.
The fact that similar fractions of 2G+1G systems are inferred -- regardless of what compact binary parameter (i.e. effective spin, mass ratio, or primary spin magnitudes) is used to identify them -- strengthens the evidence that these independently-discovered subpopulations are indeed the same as one another and represent 2G+1G mergers.

The inferred fraction of 2G+1G mergers is high when marginalized over the full redshift range in which we define our model.
Taken at face value, this may lead to the conclusion that all observed compact binaries can be sourced from clusters, as $\mathcal{O}(10\%)$ of all cluster mergers should be hierarchical~\citep{2019PhRvD.100d3027R}.
However, we find that the fraction of hierarchical mergers depends strongly on redshift, with low fractions in the local Universe.
Thus, the local ($z\lesssim0.3$) merger rate is difficult to explain with cluster mergers alone, and further study into the redshift-dependent mass function of star clusters is necessary to compare with the observed \ac{GW} population.

\section{The hierarchical merger rate evolves faster with redshift than the rest of the population}
\label{sec:redshift}
The merger rate of the $\chi_1\approx0.7$ subpopulation may evolve more steeply than that of the main population.
Figure~\ref{fig:redshift} shows the merger rate evolution of both subpopulations as a function of redshift, as well as the posteriors on the power law index of each population's redshift distributions, $\kappa_{\rm main}$ and $\kappa_{\rm sub}$. 
The power law index of the total population, as inferred by \citet{2025arXiv250818083T}, is $3.2^{+0.94}_{-1.00}$, which lies between the values we infer for our main and subpopulations. 
The posterior on $\kappa_{\rm main}$ is less than that on $\kappa_{\rm sub}$ at \result{$98.0\%$} credibility\footnote{Studies on earlier GW catalogs \citep{2024arXiv240601679P} and with models that require distinct subpopulations to be cleanly separated in mass \citep{2025arXiv250915646B} do not find evidence for differing redshift distributions between high- and low-spin subpopulations. Thus, all 127 new BBHs introduced in \ac{GWTC-4.0} may contribute to this finding, regardless of their masses.}. 
There are therefore hints, but not yet definitive evidence, that hierarchical mergers represent a higher fraction of the \ac{BBH} population at $z = 1$ compared to $z = 0$.
If it is a real feature of the population, this result can explain the previously observed broadening of the effective spin  distribution with redshift \citep{2022ApJ...932L..19B,2025arXiv250818083T}, as we discuss in Section~\ref{sec:correlation}.

Hierarchical mergers necessarily proceed the first 1G mergers in a cluster, and therefore occur at similar or slightly later times in a clusters' history than 1G mergers do.
This expectation appears to be in contradiction with the results shown in Figure~\ref{fig:redshift}, implying either that our subpopulation is not a set of 2G+1G mergers (in contrast to our spin and mass ratio results), or that a more complex picture of star cluster contributions to the BBH merger rate is necessary to explain the data.
Assuming a cluster mass function that scales inversely with the square of the cluster mass, the metallicity-specific star formation history from \citet{2017ApJ...840...39M}, and the catalog of \texttt{Cluster Monte Carlo} simulations from \citet{2020ApJS..247...48K}, we infer the fraction of all BBHs that originated in star clusters 
to be \result{$0.18^{+0.30}_{-0.14}$ at $z=0.1$ and $0.74^{+0.89}_{-0.55}$ at $z=1$}.
This directly follows from our measured fraction of 2G+1G mergers in the total population at different redshifts, which is \result{$0.03^{+0.05}_{-0.02}$} at $z=0.1$ and \result{$0.09^{+0.11}_{-0.07}$} at $z=1$.

Thus, if our assumptions about the redshift-dependent mass function of clusters are accurate, contributions from other formation scenarios are necessary to explain the merger rate at low redshift, but clusters can explain the full merger rate at high redshift.
Additionally, the fact that the fraction of cluster-origin BBHs at $z=1$ has support above unity means that denser clusters at high redshift may be required to explain our inferred fraction of 2G+1G mergers. 
Theoretical uncertainties on cluster properties as a function of redshift remain uncertain, and future work will aim to constrain these with the subpopulation's inferred redshift distribution~\citep[e.g.][]{2023MNRAS.522.5546F}.
We further discuss the potential astrophysical implications of our inferred redshift distributions in Section~\ref{sec:discussion}.

\section{Mass distributions}
\label{sec:mass}
\begin{figure*}
    \centering
    \includegraphics[width=\linewidth]{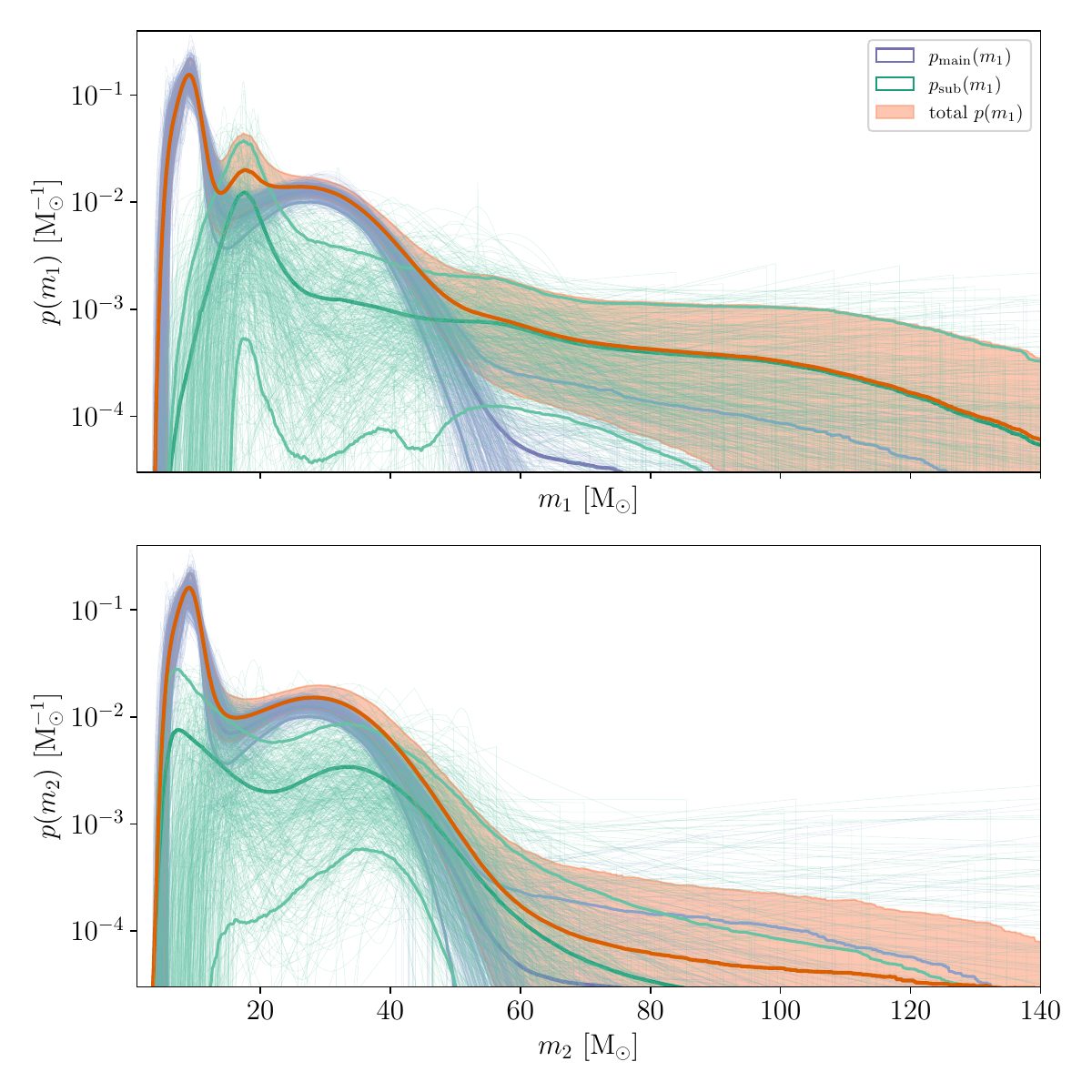}
    \caption{Inferred primary (top panel) and secondary (bottom panel) mass distributions for the hierarchically merged subpopulation (green) and main population (violet), after a pairing function is removed.
    The primary mass distribution of the subpopulation peaks at roughly twice the value of its secondary mass distribution's mode.
    Additionally, only the subpopulation's primary mass distribution contributes significantly to the merger rate at high masses.
    These two facts imply that the $\chi_1\approx0.7$ subpopulation is primarily made up of \ac{2G}+\ac{1G} mergers, whereas \ac{1G}+\ac{1G} mergers (originating either in star clusters or in the galactic field) constitute the main population.
    It is therefore possible to interpret the subpopulation's primary and secondary mass distributions as the distributions of \ac{2G} and \ac{1G} \acp{BH} in clusters, respectively.
    }
    \label{fig:mass}
\end{figure*}
Our inferred component mass distributions for each subpopulation are shown in Fig.~\ref{fig:mass}. 
As in \citet{2024ApJ...962...69F}, we display the underlying component mass distributions (i.e. after the pairing function has been factored out), but we provide marginalized component mass distributions in Appendix~\ref{ap:additional plots}.
The hierarchical subpopulation's primary mass distribution (representing 2G \acp{BH}) has a mode at \result{$17.0^{+18.3}_{-4.4}\Msun$}, which may be approximately twice as large as the mode of its secondary mass distribution (\result{$10.5^{+29.7}_{-4.7}\Msun$}).
This is consistent with a 2G+1G interpretation of the subpopulation.
    
Additionally, the subpopulation contributes the majority of the merger rate at high primary masses ($\gtrsim60\Msun$), as shown by the agreement between the orange (total) and green (subpopulation) bands in that region.
The subpopulation's secondary mass distribution does not extend to high masses, whereas the main population's primary and secondary mass distributions appear to be similar. 
Even though their maximum masses are allowed to differ, the primary and secondary mass distributions of the main (1G+1G) population both exhibit negligible merger rates above \result{$\approx 46 \Msun$} (see Fig.~\ref{fig:m99} in Appendix~\ref{ap:additional plots}).
Taken together, these results further support the hypothesis proposed in \citet{2025arXiv250904151T} that the pair-instability mass gap is polluted primarily by 2G+1G mergers rather than by 2G+2G mergers.
However, it is possible that \ac{2G}+\ac{2G} mergers are present but not yet distinguishable from the rest of the population, and we cannot rule out their presence.

The subpopulation's primary mass distribution need not have a secondary mode at higher masses, whereas the main population's primary mass distribution is clearly bimodal, with a secondary mode at $\approx35\Msun$.
This creates a ``shelf'' rather than a peak in the total primary mass distribution, which has been observed by \citet{2025arXiv250818083T}.

Our inferred mass distributions are qualitatively consistent with those found by previous studies that have searched for correlations between mass and spin \citep{2022ApJ...932L..19B,2024arXiv240601679P,2025arXiv251025579T,2025arXiv251105316T,2025arXiv250915646B,2025arXiv251203152B}.

\begin{figure}
    \centering
    \includegraphics[width=\linewidth]{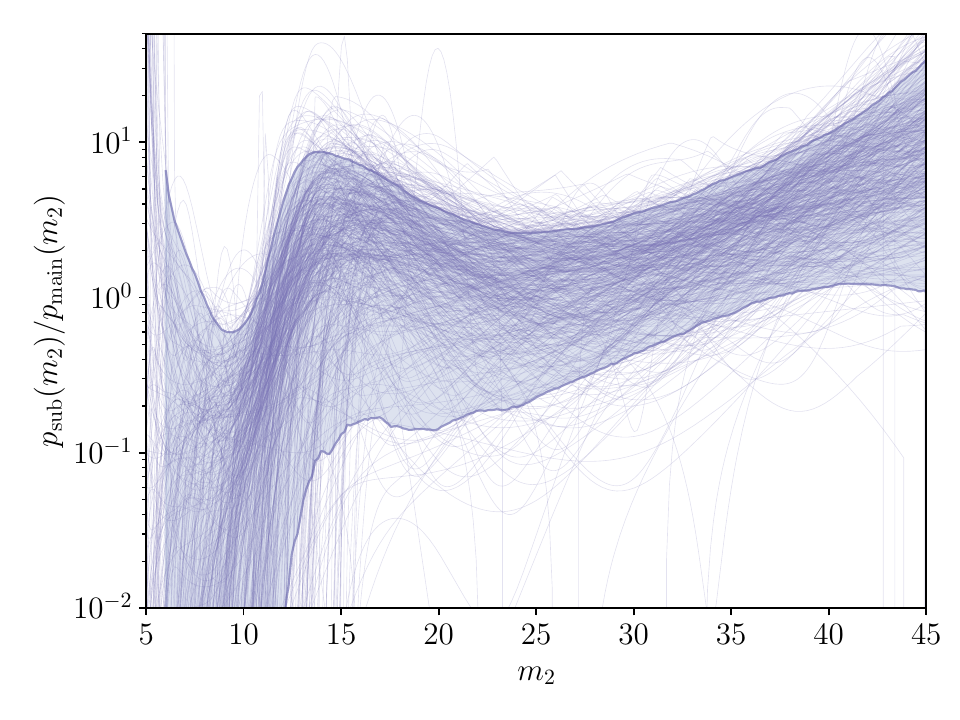}
    \caption{Ratio of the subpopulation's \ac{1G} mass distribution, representing \ac{1G} BHs that merge in clusters, to the main population's mass distribution, which consists of \ac{1G} BHs in clusters as well as all other binary formation scenarios.
    Horizontal lines on this plot would indicate that all \acp{BBH} originate in the same population of star clusters that produce 2G+1G mergers.
    However, a noticeable dip is present at $\approx9\Msun$, implying either that isolated binary evolution is required to fully explain the feature there, or that the clusters that produce our observed 1G+1G mergers differ from those that produce the 2G+1G population.}
    \label{fig:m2 comparison}
\end{figure}

A unique feature of our method is its ability to isolate the mass distribution of 1G BHs in clusters, as this is a natural interpretation of the subpopulation's \textit{secondary} mass distribution.
The low-mass peak is suppressed in this distribution relative to that of the main population.
This is especially clear in Fig.~\ref{fig:m2 comparison}, which shows the ratio of the subpopulation's $m_2$ distribution to the main population's $m_2$ distribution, revealing a noticeable dip between $8$--$10\Msun$.

This implies that the $9\Msun$ peak cannot be entirely explained by the dynamical formation hypothesis \citep[as previously proposed by ][]{2023arXiv230401288G,2022ApJ...940..184V}, unless the \ac{1G} \ac{BH} population in star clusters differs significantly between those that participate in \ac{1G}+\ac{1G} mergers versus \ac{2G}+\ac{1G} mergers.
This difference goes beyond what would already be produced by a mass ratio-dependent pairing function, which naturally favors more equal-mass pairings (and therefore selects more massive \ac{1G} \acp{BH} to participate in \ac{2G}+\ac{1G} mergers), as Figs. \ref{fig:mass} and \ref{fig:m2 comparison} use the component mass distributions with the pairing function factored out.
Such a pairing function should account for mass segregation\footnote{It is possible that our mass ratio-dependent pairing function is an insufficient descriptor of BBH pairing in clusters. For example, if BBH pairing depends on non-mass-ratio paramters, or if the pairing mechanism differs between 2G+1G versus 1G+1G mergers, as might be expected if BH ejections and other cluster dynamics cause the pairing function to change over the clusters' histories.}.
Therefore, isolated evolution may be needed to explain the low-mass end of the observed population.
This is consistent with the spin orientation distribution of the main \ac{1G}+\ac{1G} population, which has an aligned-spin component (see Appendix~\ref{sec:spin tilt}), and may therefore require a contribution from active galactic nuclei or isolated binary evolution.

\section{Hierarchical Mergers Can Source Correlations Between Effective Spin, Redshift, Mass, and Mass Ratio}
\label{sec:correlation}
\begin{figure*}
    \centering
    \includegraphics[width=\linewidth]{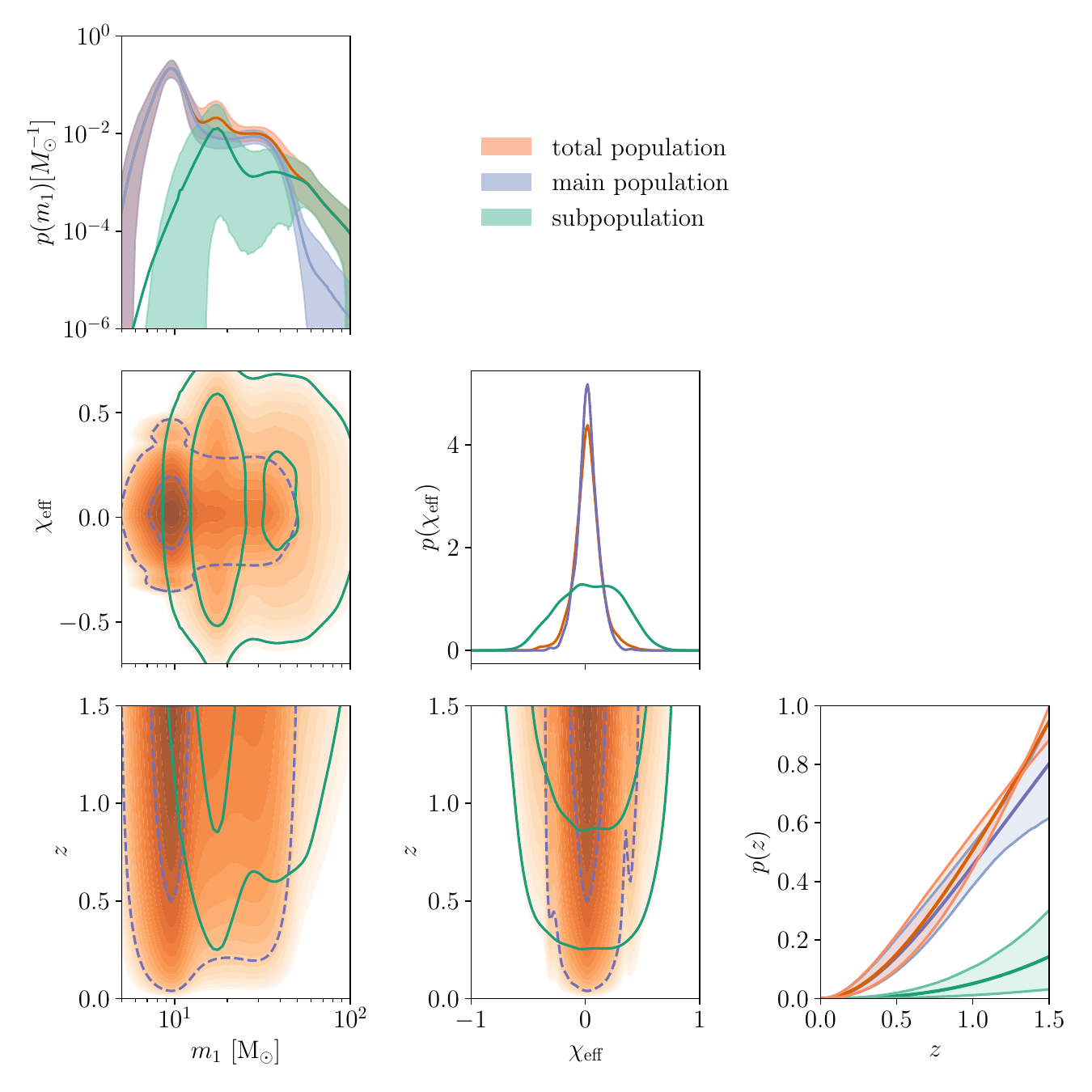}
    \caption{Population-level correlations caused by hierarchical mergers.
    Diagonal panels show the inferred one-dimensional mass, effective spin, and redshift distributions of the main population (violet), 2G+1G subpopulation (green) and combined, or total population (orange).
    These are marginalized over all other dimensions.
    Off-diagonal panels show filled contours of the total population in orange, with locations of the main and subpopulation contributions indicated with unfilled dashed and solid contours, respectively.
    All two-dimensional contours show hyperposterior-averaged distributions, while the one-dimensional mass and redshift distributions show hyperposterior uncertainty as shaded bands.
    Correlations can be observed between all parameters shown: the high-mass tail of the population increases with redshift, the effective spin distribution broadens with redshift, and the effective spin distribution modulates with mass.
    }
    \label{fig:population corner}
\end{figure*}
Fig.~\ref{fig:population corner} summarizes the results of our analysis in the context of population-level correlations. 
Note that the effective spin ($\chi_{\rm eff}$) distribution cannot be analytically calculated from our component spin model, so it is reconstructed from samples and is therefore shown as a histogram, marginalized over hyperparameter uncertainty. 
Additionally, the redshift distribution displayed in Fig.~\ref{fig:population corner} (bottom-right panel) is the (normalized) probability distribution over redshift, $p(z)$, -- as opposed to the volumetric merger rate shown in Fig.~\ref{fig:redshift} -- because $p(z)$ is directly obtained 
by marginalizing the two-dimensional probability distributions shown in the off-diagonal panels.

All pairs of parameters shown in Fig.~\ref{fig:population corner} exhibit correlations in the total population (orange contours), even though the correlations are absent in the individual subpopulations.
In particular, the hierarchical subpopulation's high primary spin magnitudes and isotropic spin tilts (see Appendix~\ref{sec:spin}) cause it to have a broader $\chi_{\rm eff}$ distribution than our main population (center panel).
Its low contribution to the merger rate at low redshift but steep increase in contribution at high redshift therefore causes the overall-$\chi_{\rm eff}$ distribution to broaden with redshift (bottom center panel), offering an explanation to the correlation first discovered by \citet{2022ApJ...932L..19B}.

In a companion paper~\citep{2026arXiv260103457V}, we show that a subpopulation of hierarchical mergers also offers a natural explanation of the correlation first discovered in \citet{2021ApJ...922L...5C}, and then found by \citet{2025arXiv250818083T} to be a narrowing of the effective spin distribution with mass ratio in \ac{GWTC-4.0} \citep[but see][for an explanation from active galactic nuclei]{2025ApJ...987...65L}.
Two previously identified correlations in the GW data can therefore be attributed to a single origin.

Thus, \acp{BBH} may be suffering from inverse Simpson's paradox~\citep{simpsons_paradox}, whereby the trends observed in the aggregate data (i.e. correlations between effective spin, redshift, and mass ratio in the full \ac{BBH} population) disappear or reverse when the data is considered in separate groups (in this case, 2G+1G versus 1G+1G BBHs).

An important caveat to our interpretation is that if the broadening of the $\chi_{\rm eff}$ distribution with redshift were not caused by hierarchical mergers, our model might spuriously identify it as a steeper redshift dependence of a highly-spinning subpopulation relative to the main population. 
One proposed explanation is the correlation between tidal spin-up and delay-time in binary star systems, as both depend on orbital separation \citep{2022A&A...665A..59B}, but we disfavor this explanation as our subpopulation appears to have an isotropic spin tilt distribution.
Regardless of the interpretation, however, our findings reinforce the fact that spin is correlated with redshift on a population level.

\section{Discussion}
\label{sec:discussion}
Using the prediction that second-generation BHs should have dimensionless spin magnitudes of $\chi\approx0.69$, we identify a subpopulation of hierarchical mergers in \ac{GWTC-4.0}.
The merger rate and primary mass distribution of this subpopulation is consistent with those found previously with alternative methods \citep{2021ApJ...913L..19T, 2022PhRvD.106j3013M, 2022ApJ...941L..39W, 2024arXiv240601679P,2025PhRvL.134a1401A, 2025arXiv251025579T, 2025arXiv251105316T, 2026arXiv260103457V}.
Having isolated this hierarchically-formed subpopulation, we study its component mass, spin orientation, and redshift distributions.

We find that the rate of hierarchical mergers likely evolves more steeply with redshift than the rest of the population.
This would imply that 2G+1G mergers occur at earlier times than many of the the 1G+1G mergers observable in current \ac{GW} data.
This is potentially surprising, as hierarchical mergers cannot occur before \ac{2G} \acp{BH} are made through \ac{1G}+\ac{1G} mergers. 
\citet{2024ApJ...967...62Y} shows that if the cluster mass function does not evolve with redshift, no significant difference is expected between the redshift evolution of 1G+1G versus hierarchical mergers (except at very early times before any \ac{2G} \acp{BH} have formed).
Our results that hierarchical mergers tend to occur at \emph{higher} redshifts than \ac{1G}+\ac{1G} mergers, and that the fraction of hierarchical mergers may be as high as \result{$0.09^{+0.11}_{-0.07}$} at $z = 1$, may therefore imply that the cluster mass function does in fact evolve with redshift, and that hierarchical mergers originate in star clusters that are denser, more massive, and at higher redshift than the lower-redshift population that hosts the rest of the observed mergers~\citep[e.g.][]{2026ApJ...998..138M}.
This interpretation is consistent with recent near-infrared observations, which suggest that high-redshift clusters are formed with higher densities, masses, and occurrence rates than the low-redshift clusters against which many cluster simulation codes are calibrated \citep{2022ApJ...940L..53V,2022ApJ...937L..35M,2024Natur.632..513A,2024Natur.636..332M,2025ApJ...981L..28M}.
Furthermore, cluster mass may increase with decreasing metallicity, which would also cause higher-mass clusters at higher redshifts.

We do not expect redshift evolution of other cluster properties, such as compactness or metallicity, to explain our results, as their effect on the ratio of 2G+1G to 1G+1G mergers is subdominant to the cluster mass.
While preliminary investigations on the \texttt{Cluster Monte Carlo} catalog confirm these arguments, detailed study on these properties is necessary and will be the focus of future work.


Another potential explanation for the steeper redshift evolution of a hierarchical subpopulation is that isolated binary stars source the observed low-redshift mergers.
These systems may have longer delay times between star formation and merger, or may have been formed later in the Universe's history than stars born in clusters.
The latter is a natural consequence of the fact that the star formation rate in clusters peaks at higher redshifts than the total star formation rate, even when accounting for the low metallicities that source massive \acp{BH}.
This hypothesis may be favored if the $9\Msun$ peak has a shallower redshift evolution than the rest of the mass distribution, as Fig.~\ref{fig:m2 comparison} suggests that non-star cluster channels contribute to this feature.
This would be in line with previous studies focusing on the spin distribution \citep{2023arXiv230401288G} and precise location \citep{2022ApJ...940..184V} of the $9\Msun$ peak.

The steeper redshift evolution of the 2G+1G population when compared to the 1G+1G population offers an explanation for the effective spin distribution's broadening with redshift~\citep{2022ApJ...932L..19B}. 
If all observed correlations in the \ac{GW} data can be attributed to a subpopulation of hierarchical mergers \citep[see e.g.][]{2026arXiv260103457V}, a consistent picture of the \ac{BBH} population is beginning to emerge.
This allows us to begin interpreting the rest of the \ac{BBH} population, using the 2G+1G subpopulation as an anchor.

Modeling choices add an important caveat to the results presented here.
While we simultaneously fit all hyperparameters of the redshift, spin tilt, and mass distributions for the main and subpopulations, we have employed a strongly-parametrized approach.
Thus, our population model has some inflexibility in the morphologies it can produce.
This would be problematic if features of the true 1G+1G population that cannot be captured by our model for the main population were erroneously absorbed into the subpopulation. 
For example, if the true 1G+1G population has a high-spin magnitude tail that cannot be accurately modeled by a truncated normal distribution, or if it has differing secondary versus primary mass distributions caused by binary interactions \citep[as in][]{2024ApJ...962...69F}, the subpopulation may absorb those residual effects.
If such effects correlate with redshift, they could contribute to the trend observed in Fig.~\ref{fig:redshift}, causing a misattribution of this trend to a hierarchically merged subpopulation. 
A useful improvement to our method would therefore be the use of hybrid strongly- and weakly-parametrized approaches, such as the one presented in \cite{2023arXiv230401288G}.

The population properties of hierarchical mergers inferred in this work will enable constraints on the properties of star clusters from which they are sourced.
For instance, interpreting the secondary mass distribution of the 2G+1G subpopulation as the birth mass distribution of BHs in star clusters can inform the metallicity content of these clusters, as well as the supernova mechanism. 
Additionally, the the 2G+1G subpopulation's redshift distribution informs the formation times, metallicity content, and redshift-dependent mass function of star clusters.

\begin{acknowledgements}
The authors gratefully acknowledge helpful conversations with Hui Tong, Daniel Holz, Utkarsh Mali, Mike Zevin, Sharan Banagiri, and Thomas Dent.
We extend special thanks to Bart Ripperda for granting us access to the \textsc{Bee} workstation at CITA, where the computations for this work were performed.
AV acknowledges support from the Natural Sciences and Engineering Research Council of Canada (NSERC) (funding reference number 568580).
MF acknowledges support from the Natural Sciences and Engineering Research Council of Canada (NSERC) under grant RGPIN-2023-05511, the Alfred P. Sloan Foundation, and the Ontario Early Researcher Award.
This material is based upon work supported by NSF’s
LIGO Laboratory which is a major facility fully funded
by the National Science Foundation.
This research was supported in part by grant NSF PHY-2309135 to the Kavli Institute for Theoretical Physics (KITP).
This research has made use of the Astrophysics Data System, funded by NASA under Cooperative Agreement 80NSSC21M00561.
\end{acknowledgements}

\software{\texttt{gwpopulation} \citep{2025JOSS...10.7753T},  
          \texttt{gwpopulation\_pipe} \citep{2021zndo...5654673T}, 
          \texttt{Dynesty} \citep{2020MNRAS.493.3132S},
          \texttt{Jupyter} \citep{2007CSE.....9c..21P,kluyver2016jupyter}, \texttt{matplotlib} \citep{Hunter:2007}, \texttt{numpy} \citep{numpy}, \texttt{pandas} \citep{mckinney-proc-scipy-2010,pandas_17992932}, \texttt{python} \citep{python}, \texttt{scipy} \citep{2020SciPy-NMeth,scipy_17873309}, \texttt{Bilby} \citep{bilby_paper,bilby_paper_2,Bilby_17533961}, \texttt{corner.py} \citep{corner-Foreman-Mackey-2016,corner.py_14209694}, \texttt{Cython} \citep{cython:2011}, \texttt{JAX} \citep{jax2018github}, and \texttt{tqdm} \citep{tqdm_14231923}.
          }

\bibliography{references,NonADSReferences}
\bibliographystyle{aasjournalv7}

\appendix
\section{Population model details}
\label{ap:model}
The functional forms for the main population and subpopulation are similar, though all parameters of these models differ between the two.
Following \citet{2025arXiv250818083T}, we model the spin tilt distribution for both populations as a mixture between an isotropically-distributed component and an aligned-spin component.
Within each population, the component spin tilts -- $\cos{\theta_1}$ and $\cos{\theta_2}$, are identically and independently distributed.
The main population is required to have independent but identically distributed spin magnitudes ($\mu^{\chi_1}_M=\mu^{\chi_2}_M=\mu^\chi_M$ and $\sigma^{\chi_1}_M=\sigma^{\chi_2}_M=\sigma^\chi_M$), whereas the subpopulation is allowed to have a different distribution of primary and secondary spin magnitudes ($\mu^{\chi_1}_S\neq\mu^{\chi_2}_S$, $\sigma^{\chi_1}_S\neq\sigma^{\chi_2}_S$), to allow for the 2G+1G scenario.
We additionally fix $\mu^{\chi_1}_S=0.69$ and $\sigma_S^{\chi_1}=0.1$ to encourage the subpopulation to represent binaries with 2G primary components.
Although this is a strong prior choice, we do not enforce the existence of this subpopulation, as $\xi$ is a free parameter which is allowed to be $0$.

The mass distribution for both populations uses a pairing function formalism \citep{2020ApJ...891L..27F} with a Gaussian pairing function in mass ratio, $q=m_2/m_1$, truncated in the range $(0,1]$.
Inspired by \citet{2025arXiv250904151T} and \citet{2024ApJ...962...69F}, the hyperparameters governing both components are identical for the main population, except that we allow differing maximum masses for the primary and secondary components.
For the subpopulation, all hyperparameters are allowed to differ between the primary and secondary masses, as we expect 1G BHs to have a different mass distribution than 2G BHs.
All component mass distributions are parametrized by the \ac{bp2p} primary mass distribution from \citet{2025arXiv250818083T}.
For a given subpopulation, the resulting two-dimensional mass distribution is then
\begin{equation}
    p_j(m_1,m_2|\lambda_j) \propto \text{BP2P}(m_1|\lambda_j^{m_1})\text{BP2P}(m_2|\lambda_j^{m_2})\mathcal{N}(q|\mu^q_j,\sigma^q_j) ,
\end{equation}
where $j = [S,M]$ denotes the particular subpopulation. 
For a description of hyperparameters governing the \ac{bp2p} model, we refer readers to Appendix B.3 of \citet{2025arXiv250818083T}.

We require that the pairing function be identical for the main and subpopulations, which encodes our assumption that the 1G+1G mergers in our sample are subject to the same gravitational potential and therefore undergo the same process of mass segregation as 2G+1G mergers. 
However, this assumption might not be valid as star cluster potentials may be time-varying, or, alternatively, our observed 2G+1G mergers may come from a distinct cluster population from our observed 1G+1G mergers.
We therefore relax this assumption in Appendix~\ref{ap:alternative pairing}, and find that our main conclusions are largely unchanged, although our mass ratio distributions differ noticeably with different pairing function assumptions.

Both subpopulations have redshift distributions of the form \citep{2018ApJ...863L..41F}
\begin{equation}
    p_j(z|\kappa_j) \propto \frac{{\rm d} V_c}{{\rm d} z} (1 +z)^{\kappa_j -1} .
\end{equation}

\section{GW231123}
\label{ap:231123}
Including GW231123 does not qualitatively impact our redshift or spin results, but it does lessen the statistical significance of the finding that the subpopulation has a steeper redshift evolution than the main population.
Because GW231123 has high primary spins, it appears to be fully absorbed in the subpopulation.
It's redshift is also relatively low ($z\approx0.4$), which causes the subpopulation's inferred redshift distribution to flatten.
Thus, we find $\kappa_M<\kappa_S$ to \result{$88\%$} when GW231123 is included in the analysis, compared to \result{98\%} when it is excluded.

Furthermore, including GW231123 in the analysis impacts our inferred secondary mass distribution.
When included, the event causes the maximum secondary mass of the subpopulation to shift to large values that are inconsistent with those inferred when GW231123 is excluded. 
We therefore deem GW231123 an outlier in $m_2$ and exclude it from the analysis in the main text of this paper.
One possible interpretation of GW231123's inconsistency with the inferred secondary mass distribution of the subpopulation is that it is not a 2G+1G merger.
However, we refrain from making definitive or quantitative statements on this event.

\section{Spin Distributions}
\label{sec:spin}
\subsection{Spin Orientations}
\label{sec:spin tilt}
\begin{figure}
    \centering
    \includegraphics[width=0.5\linewidth]{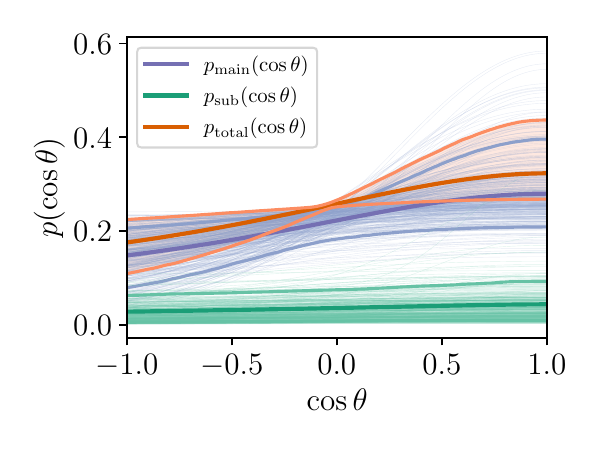}
    \caption{Distribution of spin orientations for the main population and subpopulation.
    The subpopulation prefers isotropically-oriented spins, which is consistent with expectations for a cluster origin and further validates the interpretation of the subpopulation as representing 2G+1G systems.
    The main population has support for both an isotropic component and one peaked near aligned spins.
    }
    \label{fig:spin tilt}
\end{figure}
Fig.~\ref{fig:spin tilt} shows the inferred spin orientation distributions for our main and subpopulation. 
Parameterized by $\cos{\theta_i}$, and assumed to be independent and identically distributed between the two \acp{BH} in each system, these distributions will be uniform if the spins of each BH are isotropically oriented with respect to their orbits' angular momentum and peaked near unity if they tend to be aligned with the orbital angular momentum of the system.
Isolated formation channels typically predict a preference for aligned-spin systems, while star clusters predict isotropic orientations.
Thus, the fact that the subpopulation exhibits a uniform spin orientation distribution is consistent with expectations from hierarchical mergers in dynamical environments. 

The main population, however, has support for both  isotropic and aligned-spin contributions, consistent with findings in \citet{2025arXiv250818083T}. 
Aligned-spin systems nominally constitute \result{$62^{+24}_{-25} \%$} of the main population, as indicated by the hyperposterior on the fraction of main-population systems contained within a Gaussian distribution centered at unity. 
While it is possible to interpret this as the percentage of isolated \ac{BBH} mergers in the 1G+1G population, the hyperparameter governing the fraction of systems in the Gaussian component is typically an over-estimate of the fraction of aligned-spin systems, given that broad Gaussian distributions emulate uniform distributions.

In this work, we forego making definitive or quantitative statements from the spin orientation distribution, as GW observations have limited information on spin orientations and therefore population inference is subject to prior assumptions \citep{2022A&A...668L...2V,2025PhRvD.112h3015V}. However, the results in Fig.~\ref{fig:spin tilt} encourage the interpretation of the subpopulation as a set of 2G+1G mergers in star clusters.

\subsection{Spin Magnitudes}
\begin{figure}
    \centering
    \includegraphics[width=\linewidth]{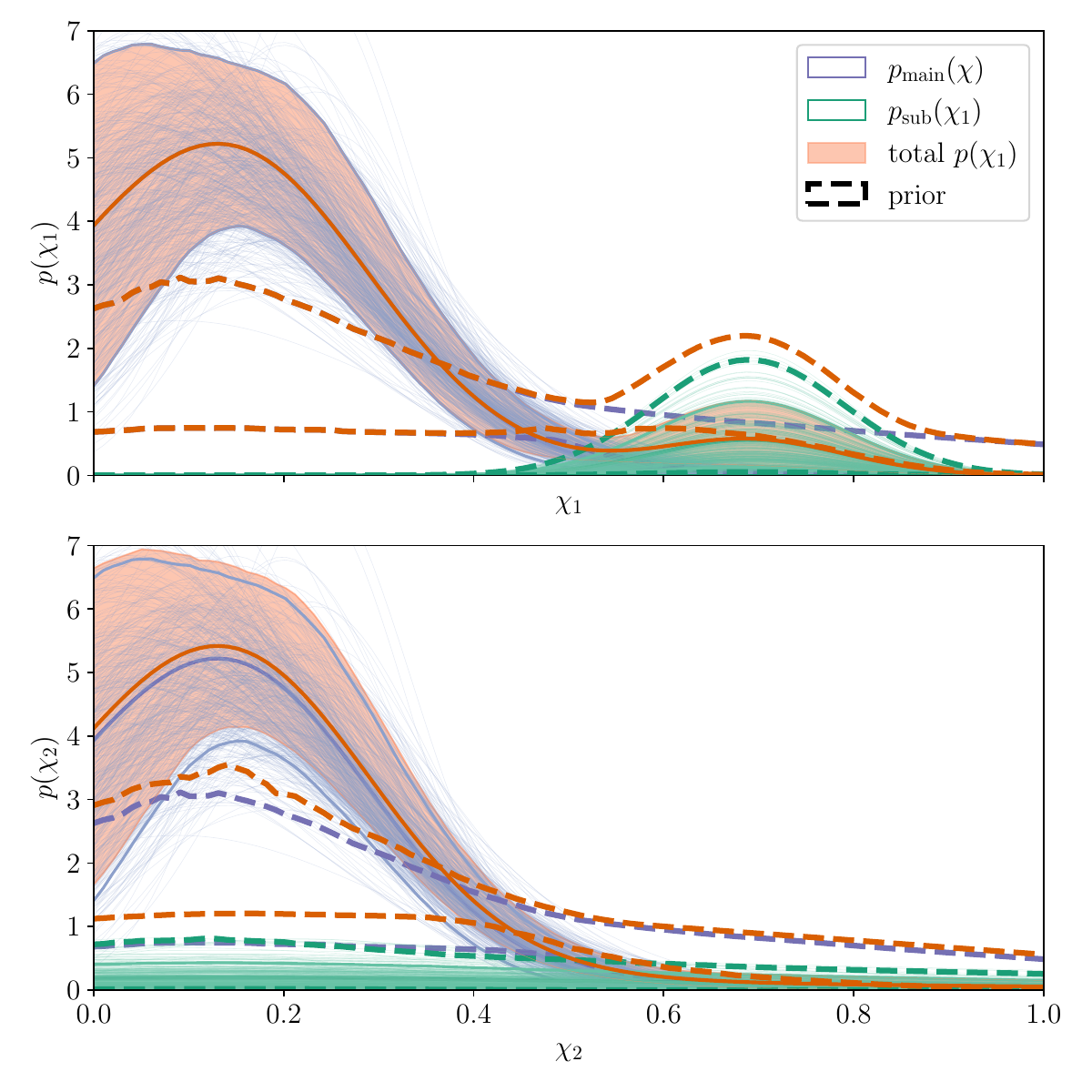}
    \caption{Distributions of spin magnitudes.
    The main population prefers low (but not necessarily zero) spins.
    The subpopulation's primary spin magnitude distribution is fixed in shape but is allowed to vary in height, whereas its secondary spin magnitude distribution resembles its prior, so the data are not yet informative to the spin magnitudes of these systems.
    }
    \label{fig:spin mag}
\end{figure}
The inferred spin magnitude distributions for our main and subpopulation are shown in Fig.~\ref{fig:spin mag}.
Remember that p($\chi_1$) for the subpopulation is a truncated Gaussian distribution with fixed mean and standard deviation, $\mu^{\chi_1}_S=0.69$, $\sigma_S^{\chi_1}=0.1$.
All uncertainty in that distribution is therefore due to the hyperposterior on $\xi$, the mixture fraction between the main and subpopulation.
The secondary spin magnitude distribution is fit to the data, and we find that it may be peaked between \result{$0.02$ and $0.23$}, indicating a preference for small but not necessarily zero spin in the 1G components.
However, the hyperposterior is broad and the spin magnitude hyperprior -- shown in dashed lines in Fig.~\ref{fig:spin mag} -- resembles the hyperposterior for the subpopulation, so we further caution against drawing conclusions from the secondary spin magnitude distribution of the subpopulation.

We assume that the two components of the main population's systems are independent and identically distributed, and find them to be peaked strongly at \result{$\mu_M^{\chi} = 0.12^{+.06}_{-.06}$}, again indicating a preference for small but nonzero spins, and consistent with \citet{2025arXiv250818083T}.

\section{Mass ratio distributions}
\label{sec:mass ratio}
\begin{figure}
    \centering
    \includegraphics[width=0.5\linewidth]{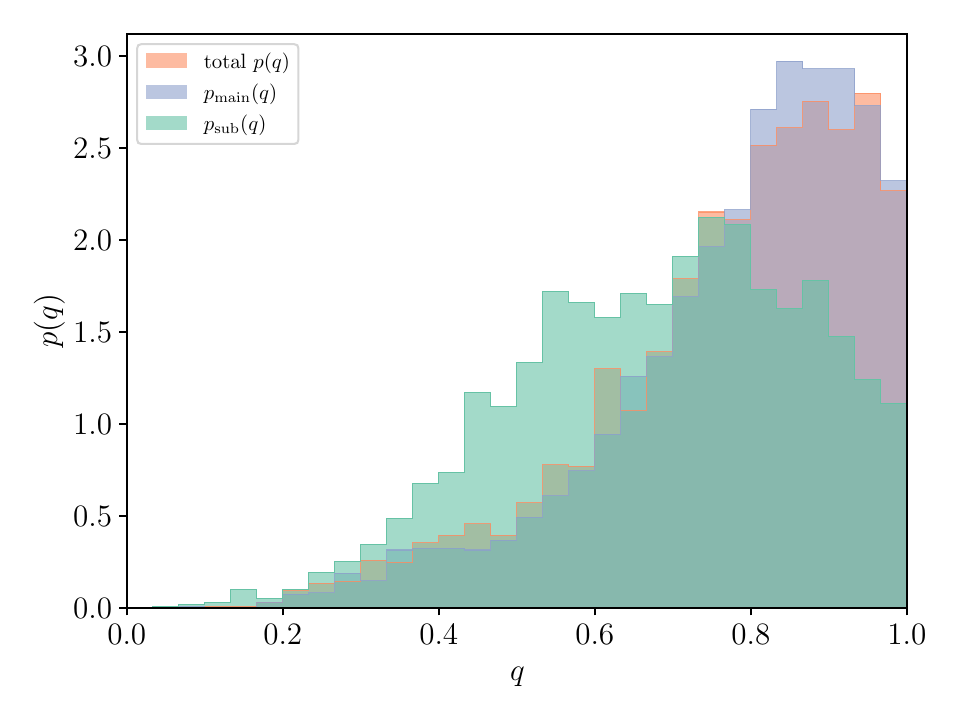}
    \caption{Mass ratio distributions, marginalized over hyperposterior uncertainty.
    The subpopulation's mass ratio distribution is peaked at lower values than that of the main population, consistent with expectations for 2G+1G mergers.
    However, the value at which it peaks depends strongly on our choice of pairing function (see Fig.~\ref{fig:pairing summary}).}
    \label{fig:mass ratio}
\end{figure}
Our inferred mass ratio distributions are influenced both by our inferred primary and secondary component mass distributions, and by our inferred pairing function.
The peak of our pairing function is inferred to be at \result{$\mu^q=0.85^{+0.07}_{-.06}$}.
This results in the mass ratio distributions shown in Fig.~\ref{fig:mass ratio}.
The subpopulation's mass ratio distribution peaks at lower values than that of the main population, consistent with expectations for 2G+1G mergers.

However, the value at which it peaks depends strongly on our choice of pairing function. 
Assuming the same pairing function between the main and subpopulation -- as we do in Fig.~\ref{fig:mass ratio} and all other figures in the main text -- results in a mass ratio distribution for the subpopulation that peaks at \result{$q=0.73$}.
Assuming a different pairing function between the main and subpopulations results in a mass ratio distribution that peaks at \result{$\approx0.5$}, which is in line with conventional expectations from 2G+1G mergers \citep{2019PhRvD.100d3027R,2026ApJ...997..267Y}.
However, the data are not yet able to constrain the pairing function of the subpopulation, so this result may be spurious.

\section{Choice of pairing function for subpopulation}
\label{ap:alternative pairing}

\begin{figure}
    \centering
    \includegraphics[width=0.5\linewidth]{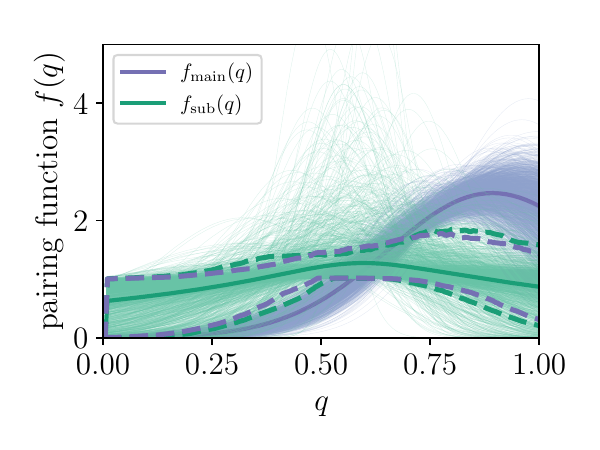}
    \caption{Inferred pairing functions for the subpopulations when these are allowed to differ. The main population's pairing function resembles what is inferred when the pairing is enforced to be identical between the two subpopulations, while the subpopulaion's pairing function resembles its prior (dashed lines).}
    \label{fig:pairing function}
\end{figure}

\begin{figure}
    \centering
    \includegraphics[width=\linewidth]{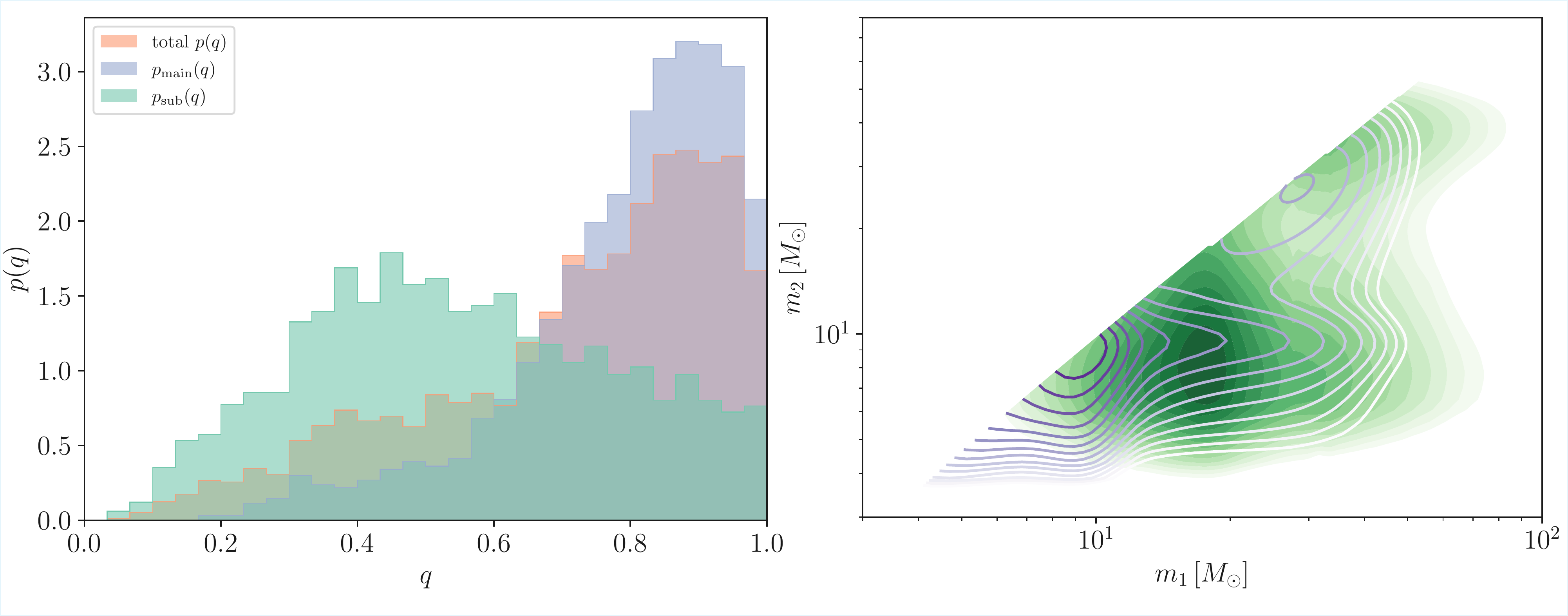}
    \caption{Mass ratio (left panel) and two-dimensional mass (right panel) distributions for the main and subpopulations in the case where the pairing function is allowed to differ between the main and subpopulation. All distributions are averaged over population-level uncertainty and colors are the same as in all previous figures. }
    \label{fig:pairing summary}
\end{figure}

The assumed form of the subpopulation's pairing function primarily affects the inferred mass and mass ratio distributions of the subpopulation. 
In the main text, we enforce that the main and subpopulation have the same pairing function in order to emulate a constant mass segregation process in star clusters.
However, this might not be a valid assumption. For example, if 2G+1G mergers happen later in a cluster's history than 1G+1G mergers, the cluster's gravitational potential and the BHs that it contains may be different by the time the majority of 2G+1G mergers happen, the mass segregation process and therefore the pairing function may be different for 1G+1G versus 2G+1G mergers.
This motivates us to explore a different pairing function for the subpopulation in this Appendix.
We show the resulting inferred pairing functions in Fig.~\ref{fig:pairing function}.
While the main population's pairing function (solid violet lines) is well-constrained and appears different from its prior (dashed violet lines), the subpopulation's pairing function resembles its prior (dashed green lines).
Furthermore, the main population's pairing function in Fig.~\ref{fig:pairing function} strongly resembles the pairing function inferred from combining the main and subpopulation.
It therefore may not be possible to meaningfully constrain the pairing of the 2G+1G population, further motivating our decision to enforce the same pairing between the main population and subpopulation in the main text. 
    
In Fig.~\ref{fig:m1m2}, we show the population-averaged mass ratio distribution and two-dimensional mass distribution for the case in which the pairing functions differ between the main population and subpopulation (compare to to Figs.~\ref{fig:mass ratio} and \ref{fig:m1m2}).
If the pairing functions differ, the subpopulation's mass ratio distribution peaks at $0.5$, which is more in line with expectations from 2G+1G mergers, though we caution that these results are likely prior-driven. 
    
Results relating to spin and redshift distributions are unchanged between the two choices of pairing functions considered here.

\section{Additional figures}
\label{ap:additional plots}
\begin{figure}
    \centering
    \includegraphics[width=\linewidth]{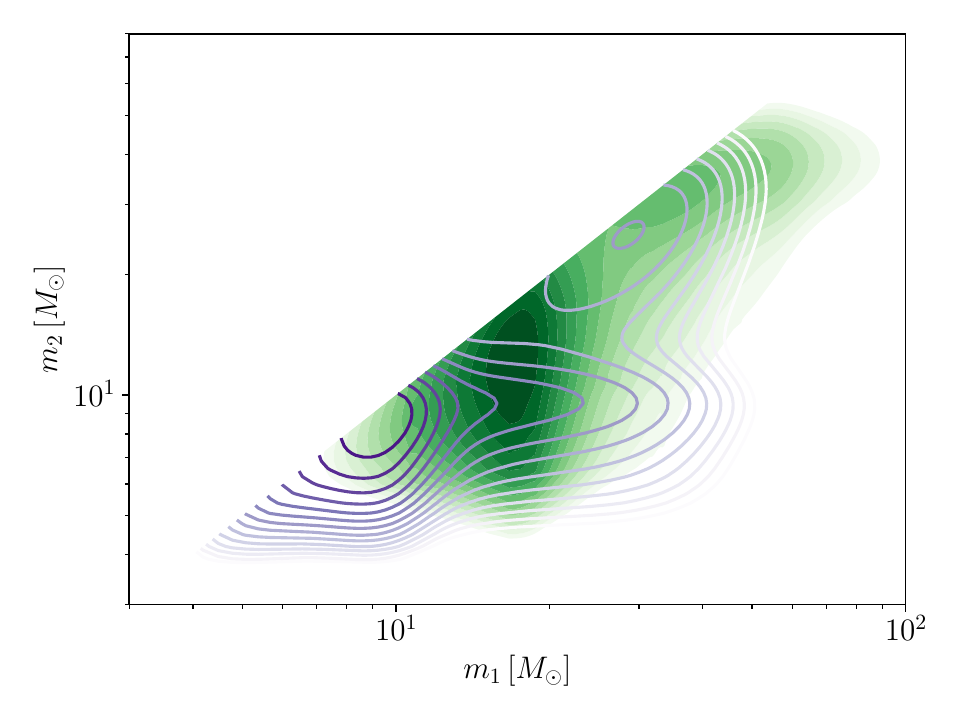}
    \caption{Two-dimensional mass distributions of the main and subpopulation. Colors are the same as in all previous figures.}
    \label{fig:m1m2}
\end{figure}
\begin{figure}
    \centering
    \includegraphics[width=\linewidth]{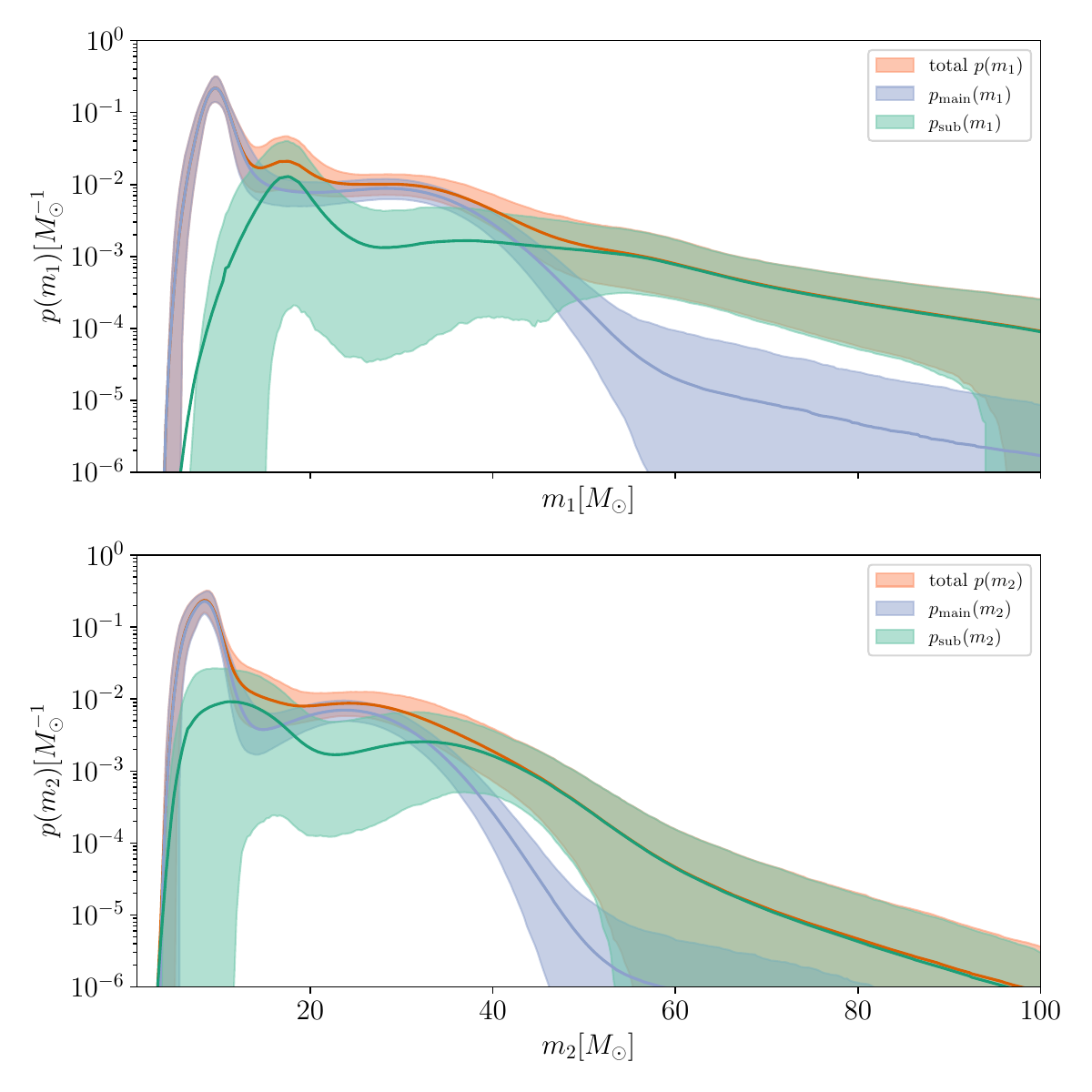}
    \caption{Marginal primary (top) and secondary (bottom) mass distributions.
    These produce similar conclusions to those presented in Section~\ref{sec:mass}.
    }
    \label{fig:marginal mass}
\end{figure}
\begin{figure}
    \centering
    \includegraphics[width=0.5\linewidth]{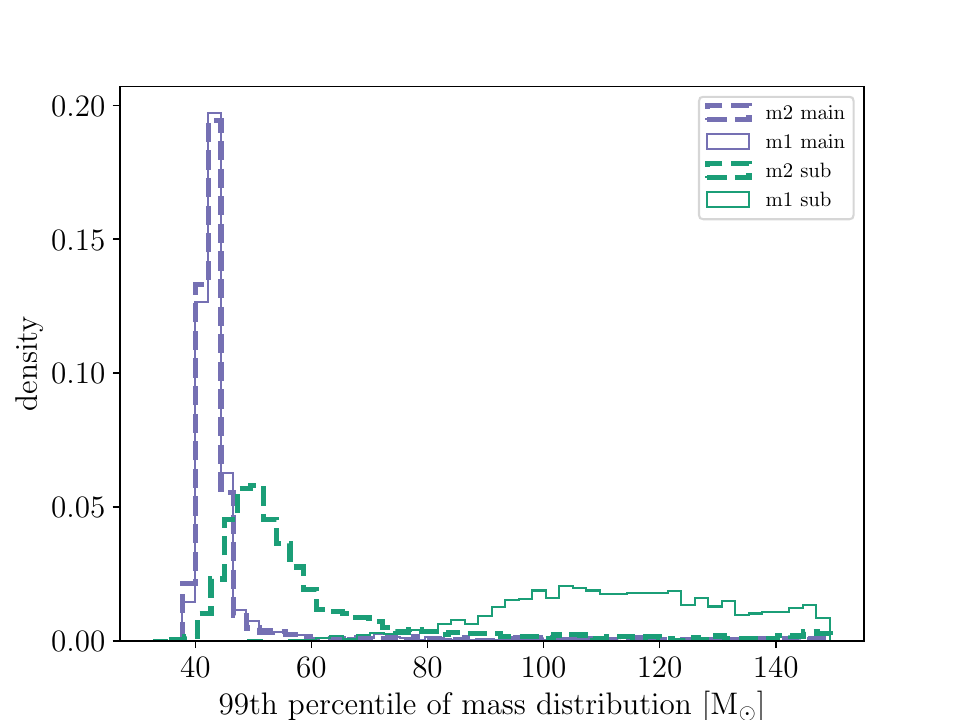}
    \caption{99th percentile of the primary (solid) and secondary (dashed) mass distributions for the main and subpopulations. These are calculated directly from Fig.~\ref{fig:mass}.
    The main population's merger rate declines above $46\Msun$ in both primary and secondary mass, whereas the subpopulation declines above $\approx55\Msun$ in secondary mass and beyond $110\Msun$ in primary mass.
    }
    \label{fig:m99}
\end{figure}
Fig~\ref{fig:m1m2} shows our inferred two-dimensional primary and secondary mass distributions of the main and subpopulations.

Fig.~\ref{fig:marginal mass} shows the marginal component mass distributions, which differ from Fig.~\ref{fig:mass} because they include the effects of the pairing function, but the same qualitative features appear in both representations.
The top panel of Fig.~\ref{fig:marginal mass} can be directly compared to Fig. 3 of \citet{2025arXiv251105316T}, and does exhibit qualitatively similar features.
Specifically, the 2G+1G population peaks at $\approx17\Msun$, may drop to negligible rates between $25\Msun$ and $30\Msun$, and does not require a secondary mode, instead preferring a shallow tail to high masses.
The 1G+1G population is also consistent between Fig.~\ref{fig:marginal mass} and Fig. 3 of \citet{2025arXiv251105316T}, with a dip at $\approx17\Msun$ and peak at $\approx9\Msun$.

Fig.~\ref{fig:m99} displays the 99th percentiles of the component mass distributions in Fig.~\ref{fig:mass}.
It would seem to imply that the pair-instability mass feature begins at $\approx46\Msun$.
This is consistent with 
\citet{2022ApJ...941L..39W}, \citet{2024arXiv240601679P}, and \citet{2025PhRvL.134a1401A}, but appears inconsistent with \citet{2025arXiv251022698W}, who find a population of low-spin objects extending to $\approx70\Msun$.
To directly compare to this finding, we calculate the fraction of systems in the spinning subpopulation in various mass bins and find that this fraction is lower in the range $m_1\in[50\Msun,70\Msun]$ than in the range $m_1\in[70\Msun,90\Msun]$ to 91.1\%.
This supports the hypothesis put forth in \citet{2025arXiv251022698W}, but at low significance.


\end{document}